\documentclass{mem}
\usepackage{natbib}\usepackage{txfonts}\usepackage{balance}
\usepackage{graphicx}
\usepackage[a4paper]{hyperref}
\idline{75}{282}
\begin{document}
\def\teff{$T\rm_{eff }$}
\def\kms{$\mathrm {km s}^{-1}$}
\def\water{H$_2$O~}

\title{
Magnetic fields around late-type stars using \water maser observations
}

   \subtitle{}

\author{
W.H.T.\ Vlemmings\inst{1}, H.J.\ van~Langevelde\inst{2} 
\and P.J.\ Diamond\inst{1}
          }

  \offprints{W.Vlemmings}

\institute{
Jodrell Bank Observatory, 
The University of Manchester,
Macclesfield,
Cheshire SK11~9DL, U.K
\email{wouter@jb.man.ac.uk}
\and
Joint Institute for VLBI in Europe,
Radiosterrenwacht Dwingeloo,
Postbus 2,
7990~AA, Dwingeloo,
the Netherlands
}

\authorrunning{W.H.T.\ Vlemmings }

\titlerunning{Magnetic fields around late-type stars}

   \abstract{ We present the analysis of the circular polarization,
due to Zeeman splitting, of the \water masers around a sample of
late-type stars to determine the magnetic fields in their
circumstellar envelopes (CSEs). The magnetic field strengths in the
\water maser regions around the Mira variable stars U~Ori and U~Her
are shown to be several Gauss while those of the supergiants S~Per,
NML~Cyg and VY~CMa are several hundred mG. We also show that large
scale magnetic fields permeate the CSE of an evolved star; the
polarization of the \water masers around VX~Sgr reveals a dipole field
structure. We shortly discuss the coupling of the magnetic field with
the stellar outflow, as such fields could possibly be the cause of
distinctly aspherical mass-loss and the resulting aspherical planetary
nebulae.  \keywords{masers -- polarization -- stars: circumstellar
matter -- stars: magnetic fields -- stars: supergiants -- stars:
Miras} }

\maketitle{}

\section{Introduction}

 The exact role of magnetic fields in the mass loss mechanism and the
formation of CSEs around late-type stars is still unclear but could be
considerable. The study of several maser species found in CSEs has
revealed important information about the strength and structure of
magnetic fields throughout the envelopes surrounding the late-type
stars. At distances from the central star of up to several thousands
of AU, measurements of the Zeeman effect on OH masers indicate
magnetic fields strengths of a few milliGauss \citep[e.g.][]{SC97,
MvL99}. Additionally, weak alignment with the CSE structure is found
\citep[e.g.][]{E04}. Observations of SiO maser polarization have shown
highly ordered magnetic fields close to the central star, at radii of
5-10 AU where the SiO maser emission occurs
\citep[e.g.][]{BM87,KD97}. When interpreting the circular polarization
of the SiO masers as standard Zeeman splitting, the magnetic field
strength determined from these observations could be up to several
tens of Gauss. However, a non-Zeeman interpretation can explain the
observations by magnetic field strengths of only several tens of
milliGauss \citep{WW98}.  Recently, high circular polarization of
circumstellar H$_2$O masers was observed for a small sample of
late-type stars \citep[][hereafter V02 and V05]{V02b, V05}.  H$_2$O
masers occur at intermediate distances in the CSE, in gas that is a
factor of $10-1000$ more dense than the gas in which OH masers
occur. Here we discuss the results of the \water maser observations
and their possible role in shaping the CSEs.

\begin{figure*}[t!]
\resizebox{\hsize}{!}{\includegraphics[clip=true]{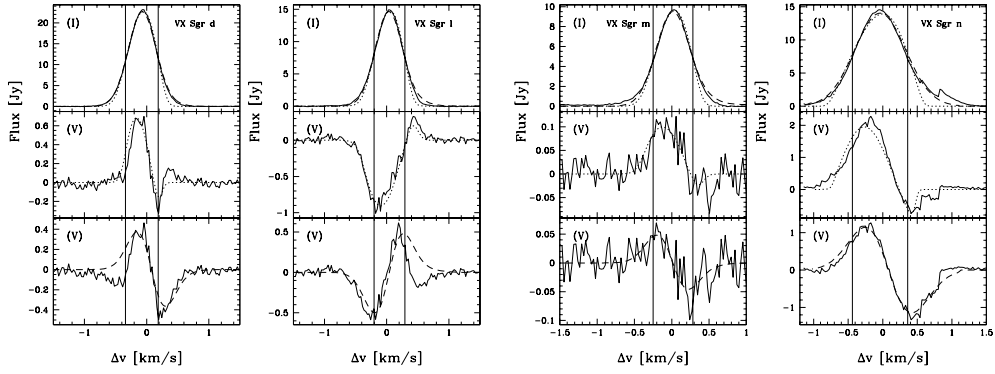}}
\caption{\footnotesize 
Examples of the total power (I) and circular
polarization (V) spectra for several of the \water maser features
around VX~Sgr. The bottom panel shows the best fitting synthetic
V-spectrum produced by the standard LTE Zeeman interpretation (dashed
line). The middle panel shows the best non-LTE model fit (dotted
line). The corresponding total power fits are shown in the top
panel. The V-spectra in the lower two panels are adjusted by removing
a scaled down version of the total power spectrum which is different
for the LTE and non-LTE fits. The solid vertical lines show the
expected position of the minimum and maximum of the V-spectrum in the
general LTE interpretation. The magnetic field strengths along the maser line of sight determined from these spectra are $-469\pm 60$, $966\pm 120$, $-167\pm 87$ and $-4082\pm 305$ respectively.}
\label{fig1}
\end{figure*}

\section{Observations}

The observations were performed at the NRAO\footnote{The National
Radio Astronomy Observatory is a facility of the National Science
Foundation (NSF) operated under cooperative agreement by Associated
Universities Inc.} Very Long Baseline Array (VLBA). The average beam
width is $\approx 0.5 \times 0.5$~mas at the frequency of the $6_{16}
- 5_{23}$ rotational transition of H$_2$O, 22.235 GHz. We used 4
baseband filters of 1 MHz width, which were overlapped to get a
velocity coverage of $\approx 44$~km/s, covering most of the velocity
range of the H$_2$O masers. The data were correlated multiple
times. The initial correlation was performed with modest ($7.8$~kHz$ =
0.1$~\kms) spectral resolution, which enabled us to generate all 4
polarization combinations (RR, LL, RL and LR). Two additional
correlator runs were performed with high spectral resolution
($1.95$~kHz$ = 0.027$~\kms) which therefore only contained the two
polarization combinations RR and LL. This was necessary to be able to
detect the signature of the H$_2$O Zeeman splitting in the circular
polarization data and to cover the entire velocity range of the \water
masers. The data analysis path is described in detail in V02.

\begin{table*}
\caption{Observed Stars}
\label{sample}
\begin{center}
\begin{tabular}{lccccccc}
\hline
\\
Star & Type & RA (J2000) & Dec (J2000) & Distance & Period & V$_{\rm rad}$ & B$_{\rm H_{2}O}$ \\
&&($^{h}~^{m}~^{s}$)&($^{\circ}~{'}~{"}$)&(pc)&(days)&(km/s)&(G)\\
\hline
\\
U~Her   & Mira & 16 25 47.4713 & +18 53 32.867 & 277 & 406 & -14.5 & $\sim 1.5$ \\
U~Ori   & Mira & 05 55 49.1689 & +20 10 30.687 & 300 & 368 & -38.1 & $\sim 3.5$ \\
VX~Sgr  & Supergiant & 18 08 04.0485 & -22 13 26.614 & 1700 & 732 & 5.3 & $\sim 0.5 - 4$ \\
S~Per   & Supergiant & 02 22 51.72 & +58 35 11.4 & 1610 & 822 & -38.1 & $\sim 0.15$ \\
NML~Cyg & Supergiant & 20 46 25.7  & +40 06 56   & 1220 & 940 & 0.0   & $\sim 0.18$ \\
VY~CMa  & Supergiant & 07 22 58.3315 & -25 46 03.174 & 1500 & 2000 & 22.0 & $\sim 0.5$ \\
\\
\hline
\end{tabular}
\end{center}
\end{table*}

\subsection{Sample} 

Our sample consists of the 2 Mira variable stars U~Her and U~Ori and
the 4 supergiants VX~Sgr, S~per, NML~Cyg and VY~CMa. Observations were
also performed on the Mira R~Cas but the \water masers were not
detected. U~Her, S~Per. VY~CMa and NML~Cyg are discussed in detail in
V02; U~Ori, VX~SGr and a second epoch of U~Her observations are
presented in V05. The stellar sample is listed in Table~\ref{sample}.

\section{Results}

The magnetic field observed in the \water maser regions of our sample
is given in table~\ref{sample}. As we only observe the component along
the line of sight at the location of the brightest maser features, the
magnetic field values are determined from the maximum of the measured
values. We determined the magnetic field strength from the observed
circular polarization using both an LTE and non-LTE analysis as
described in V02. Because the non-LTE method, using the radiative
transfer equations from \citet{NW92}, provided the best fit, we adopt
the non-LTE magnetic field strength values as the true values. The LTE
values are typically $\sim~40\%$ higher. Fig.~\ref{fig1} shows several
examples of observed total power and circular polarization spectra for
VX~Sgr.  For none of the stars in our sample were we able to detect
any significant linear polarization above a limit of $\sim 0.02\%$ for
the strongest maser features and a few \% for most of the weakest.

\subsection{Special case: VX~SGr}
 
\begin{figure*}[]
\resizebox{\hsize}{!}{\includegraphics[clip=true]{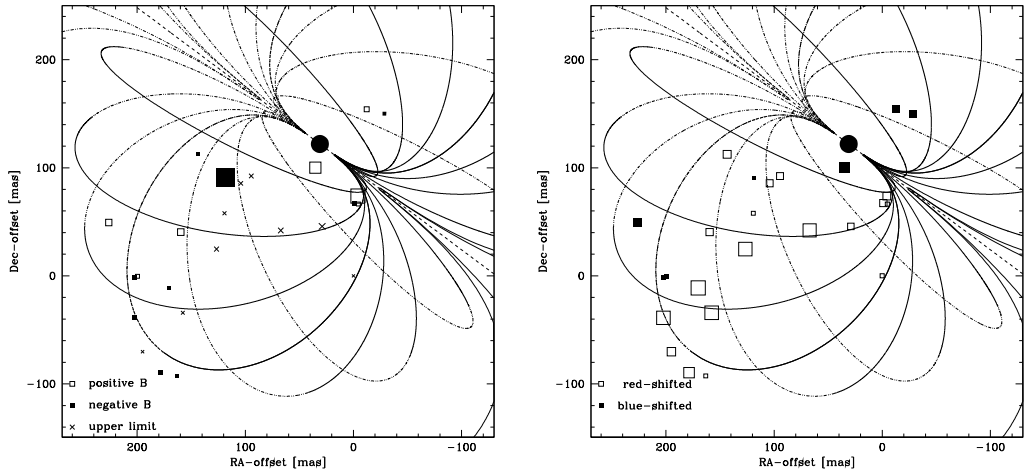}}
\caption{
\footnotesize
The best fitted dipole magnetic field for the \water
maser observations around VX~Sgr (denoted by the solid circle). (left)
The distribution of the \water maser features indicating the measured
magnetic field strengths. Open symbols denote a positive magnetic
field while the closed symbols correspond to a negative magnetic
field. The crosses represent the upper limits. The symbols have been
scaled relative to the magnetic field strength. (right) The
distribution of the \water maser features indicating the velocity
structure of the maser features. The open symbols are the red-shifted
features and the solid symbols are the blue-shifted features. The size
indicates the velocity difference with the stellar velocity ($v_{\rm
rad} = 5.3$~\kms)
}
\label{fig2}
\end{figure*}

 The rich \water maser structure around VX~Sgr for the first time
enabled us to study the morphology of the magnetic field in the maser
region in detail. The maser shell of VX~Sgr has been studied
extensively with MERLIN and VLA observations \citep[e.g.][]{L84, CC86,
ZF96, T98}. The recent observations of \citet[][hereafter M03]{MYRC03}
indicate that the \water masers arise in a thick shell between
$100$~AU and $325$~AU from the star. The \water maser shell shows a
clear elliptical asymmetry which was modelled in M03 as a spheroidal
maser distribution intersected with an under-dense bi-conical
region. In our observations we see a clear transition between a
negative magnetic field in the S-E to a positive magnetic field in the
N-W. We managed to fit a dipole magnetic field to our \water maser
magnetic field observations with the results shown in
Fig.~\ref{fig2}. The best fitted model is a dipole with its magnetic
axis pointed toward us at an inclination angle $i=40 \pm 5^\circ$, and
a position angle of $\Theta = 220 \pm 10^\circ$.  The fitted values
are remarkably consistent with the values found for the magnetic field
determined from OH masers, as well as with the orientation angle
determined from the \water maser distribution in M03 and
\citet{M96}. As a result of our fit, we find that the magnetic field
strength at the surface of VX~Sgr corresponds to $B\approx
2.0\pm0.5$~kG, consistent with the OH and SiO maser observations
\citep{BM87}. This is the first detection of a large scale magnetic
field in a circumstellar \water maser region.

\begin{figure}[h]
\resizebox{\hsize}{!}{\includegraphics[clip=true]{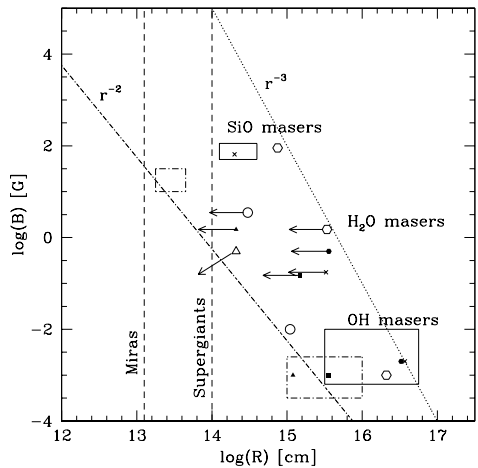}}
\caption{ \footnotesize 
Magnetic field strength, $B$, as function of
distance, $R$, from the centre of the star. Dashed-dotted boxes are
the estimates for the magnetic fields in the OH and SiO maser regions
of Mira stars, solid boxes are those for the supergiants. The
dashed-dotted line indicates a solar-type magnetic field and the
dotted line indicates the dipole field. The solid symbols are the
stars observed during our first observations (U~Her: triangles; S~Per:
squares; VY~CMa: crosses and NML~Cyg: hexagonals). The open symbols
indicate the stars of the sample observed during our second
observations (U Her: triangle; VX~Sgr: hexagonal and U~Ori:
circle). Note that the open triangle only indicates the upper limit
determined for the most recent observation of U~Her. Also note that
the magnetic field strength observed on the OH masers of U~Ori is
larger than the typically observed fields for Mira variables. The
dashed lines represent estimates of the stellar radius.  
}
\label{fig3}
\end{figure}

\section{Discussion}

 The observed magnetic field strengths on the \water masers are
consistent with the results for the other maser species assuming a $B
\propto r^\alpha$ dependence of the field strength on the distance to
the star. As the magnetic field strength determined for the SiO,
\water and OH maser features depends on the angle between the magnetic
field and the maser propagation axis, it is difficult to exactly
determine the value of $\alpha$. In Fig.~\ref{fig3} we show the
relation between the magnetic field strength and the distance from the
star for the stars in our sample. The points for the \water masers are
drawn at the outer radius of the maser region which is an upper limit.
Extrapolating the observed magnetic field strengths to the stellar
surface, we find that Mira variable stars have surface field strengths
up to several times $10^2$~G, while supergiant stars have fields of
the order of $10^3$~G. Such high magnetic field strengths indicate
that the standard Zeeman interpretation of the SiO maser polarization
is most likely correct.

 The origin of the strong, large scale magnetic fields around evolved
stars remains a topic of debate. The generation of an axisymmetric
magnetic field requires a magnetic dynamo in the interior of the
star. Several models have been discussed in the literature that
include the interaction between the differential rotation and
turbulence in the convection zone around the degenerated core
\citep[e.g.][]{BF01}. However, a supergiant core is supposedly not
degenerate.  A dynamo driven by the differential rotation between the
contracting non-degenerate core and the expanding outer envelope has
also been shown to also be able to generate strong magnetic fields
\citep{UB82}.  In alternative models, a strong magnetic field can be
generated when the star is spun up by a close binary. The sources in
our sample do not however, show any indication of binarity, although
this cannot be ruled out.

 The magnetic field strength of several Gauss measured in the \water
maser region implies that the magnetic pressure dominates the thermal
pressure of the circumstellar gas throughout a large part of the CSE
of both regular AGB stars and supergiants.  The effects of magnetic
fields on the stellar outflow and the shaping of the distinctly
aspherical PNe have been discussed in several papers
\citep[e.g.][]{P87, CL94, G-S97}. Strong, large scale, magnetic fields
around AGB stars could directly affect the fast wind that shapes the
PN or could have shaped the initial matter distribution during the
slow wind. For instance, a dipole field has been shown to be able to
create an equatorial, possibly warped, disk \citep{MB00}. Interactions
with a warped disk have been shown to be able to form multipolar PNe \citep{I03}.

\section{Conclusions}

 We have measured the strong magnetic field around a sample of
late-type stars using observations of the circular polarization of
their \water masers. Observations of VX~Sgr also indicate that the
magnetic field has a dipole shape that can be responsible for shaping
the outflow. For the lower mass late-type stars the magnetic field can
be the cause of aspherical PNe.

\bibliographystyle{aa}

\end{document}